\def\etal{{\it et al.~\/}}

\def\ie{{\it i.e.~\/}}

\def\ltsima{$\; \buildrel < \over \sim \;$}
\def\simlt{\lower.5ex\hbox{\ltsima}}
\def\gtsima{$\; \buildrel > \over \sim \;$}
\def\simgt{\lower.5ex\hbox{\gtsima}}

\documentstyle[12pt,aasms4,psfig]{article}
\tighten
\singlespace

\lefthead{Marri  \& Ferrara}
\righthead{Gravitational Magnification of Pop~III Supernovae}

\begin{document}

\title{Gravitational Magnification of Pop~III Supernovae in \\
Hierarchical Cosmological Models: NGST Perspectives}

\author{Simone Marri$^1$, Andrea Ferrara$^2$}
\affil{
$^1$Dipartimento di Astronomia, Universit\`a di Firenze, \\
50125 Firenze, Italy 
\\ E--mail: marri@arcetri.astro.it\\
$^2$Osservatorio Astrofisico di Arcetri \\ 50125 Firenze, Italy 
\\ E--mail: ferrara@arcetri.astro.it} 

\begin{abstract}
We study the gravitational lensing magnification produced
by the intervening cosmological matter distribution, as deduced from 
three different hierarchical models (SCDM, LCDM, CHDM)
on very high redshift sources, particularly supernovae in protogalactic
(Pop III) objects. By means of ray-shooting numerical simulations we find 
that  caustics are more intense and concentrated in SCDM models. 
The magnification probability function presents a moderate degree of evolution
up to $z\approx 5$ (CHDM) and $z\approx 7$ (SCDM/LCDM). All models predict that
statistically large magnifications, $\mu \simgt
20$ are achievable, with a probability of the order of a fraction of percent,
the SCDM model being the most efficient magnifier.
All cosmologies predict that above $z\approx 4$ there is a
10\% chance to get magnifications larger than 3.
We have explored the observational perspectives for Pop~III SNe detection
with NGST when gravitational magnification is taken into account.
We find that NGST should be able to detect and confirm spectroscopically 
Type II SNe up to
a redshift of $z \approx 4$ in the J band (for $T_{SN}=25000$~K);
this limit could be increased
up to $z\approx 9$ in the K band, allowing for a relatively moderate
magnification.
Possibly promising strategies to discriminate among cosmological models
using their GL magnification predictions and very high-$z$ SNe are sketched.
Finally, we outline and discuss the limitations of our study. 
\end{abstract}

\keywords{Cosmology: theory -- dark matter -- gravitational lensing --
supernovae -- galaxy evolution -- methods: numerical -- statistical}

\section{Introduction}

Current models of cosmic structure formation based on CDM scenarios
predict that the first collapsed, luminous (hereafter Pop~III) objects 
should form at redshift $z\approx 30$ and have a total mass $M \approx 
10^6 M_\odot$ or baryonic mass $M_b \approx 10^5 M_\odot$
(Couchman \& Rees 1986, Ciardi \& Ferrara 1997, Haiman \etal 1997, 
Tegmark \etal 1997, Ferrara 1998). This conclusion is reached by requiring
that the cooling time, $t_c$, of the gas is shorter than the Hubble time, $t_H$,
at the formation epoch. In a plasma of primordial composition the only efficient
coolant in the temperature range $T\le 10^4$~K, the typical virial temperature
of Pop~III dark matter halos, is represented by H$_2$ molecules whose abundance
increases from its initial post-recombination relic value to higher values during 
the halo collapse phase.

Thus, as the collapse proceeds, the gas density increases and stars are
likely to be formed. However, the final product of such star formation 
activity is presently quite unknown. This uncertainty largely depends on
our persisting ignorance on the fragmentation process and its relationship with the 
thermodynamical conditions of the gas, 
both at high $z$ and, in a less severe
manner, in present day galaxies. Ultimately, this prevents firm conclusions
on the mass spectrum of the formed stars or their IMF. This problem was already
clear more than two decades ago, as pioneering works (Silk 1977; Kashlinsky
\& Rees 1983; Palla, Salpeter \& Stahler 1983; Carr \etal 1984) could not reach
similar conclusions on the typical mass range of newly formed stars in
the first protogalactic objects. Roughly speaking, two possibilities 
can be envisaged: either (i) a limited number of Very Massive Objects (VMOs), 
single stars
with mass in the range $10^2-10^5 M_\odot$ could be formed, or (ii) a more common
stellar cluster, slightly biased towards low-mass stars (or, even, some
combination of the two involving low-mass star coalescence to form a VMO).
Apart from the fact that observational evidences     of the VMO hypothesis
are still lacking in the local universe, there are also several theoretical
problems (El Eid \etal 1983; Ober \etal 1983) that make 
VMOs as unlikely stellar prototype candidates 
in the early universe. Then, if the IMF is more of the standard type, 
questions arise about its shape and median value. A common claim in the
literature is that the absence of metals (as it is the case in a collapsing
Pop III) should shift the peak and the median of the IMF toward higher masses. 
Obviously, assessing the presence of massive stars that produce ionizing photons
and die as Type II supernovae (SNe) would be of primary importance to clarify the role
of Pop IIIs in the reionization and rehating of the universe, and, in general,
for galaxy formation.
Thus, studying the formation of the first stars has become one of the
most challenging problems in physical cosmology.
It is not the aim of this paper to investigate this aspect further;
detailed studies can be found in Gnedin \& Ostriker 1997, Haiman \& Loeb 1997,  
and Ciardi \& Ferrara 1997.
Here, we simply point out that if SNe are allowed to occur at these very 
high redshifts, they could outshine their host protogalaxy by orders of
magnitude and likely become the most distant observable sources since the 
QSO redshift distribution has an apparent cutoff beyond $z\approx 4$. 
Nevertheless, even for (future) large telescopes as NGST, VLT and Keck 
this will not be an easy task, this requiring to reach limiting magnitudes
$\approx 33$ in the near IR to observe a SN exploding at $z=10$.
Fortunately, the matter distribution in the universe behaves like a gravitational
lens that distorces, and, most importantly here, magnifies the source flux.
This magnification, as we will see, strongly enhances the detection 
probability of Pop III SNe, bringing their apparent magnitudes inside the
instrumental capability range. A different, but complementary, strategy would be
to monitor the central regions of nearby galaxy clusters (Miralda-Escud\'e \&
Rees 1997), where magnifications
high enough to allow for the detection of faint objects could be produced.
We have chosen to investigate the gravitational lensing (GL) magnification
effects of the entire cosmological mass distribution to verify if the different 
GL magnification predictions of the most attractive cosmological models
could be used as a test bench of their descriptive power of the universe.
Pushing this comparison to high redshift, as we will see, is crucial since
magnification properties deduced from hierarchical models start to differentiate 
considerably only beyond $z\approx 5$.

%
Gravitational lensing is sensitive to the cosmic mass distribution, which in turn
depends on the adopted fluctuation spectrum, to the expansion factor, and to 
cosmological distances; therefore, it represents a perfect tool for cosmological
studies. As stated above, our aim is to calculate GL effects
on high $z$ sources, as for example SNe in Pop III objects, and, in particular, 
the magnification probability as a function of the source redshift and of the
cosmological model.
In order to obtain this result it is necessary to define how structure 
is formed in the universe and a method to evaluate the statistical incidence 
of GL effects. The former problem may be approached by the linear perturbation 
theory of cosmic density fluctuations, numerical N-body simulations or by
semi-analytical methods, as, for example, the Press \& Schechter (1974,
hereafter PS) formalism. 
GL can be investigated both by numerical simulations following light propagation 
or via semi-analytical estimates.
We have chosen to solve the selected problem by using simulations
based on the so-called {\it ray-shooting} numerical scheme as far as
GL is concerned and to derive the mass and redshift distribution of the 
lens ensamble from PS. The reasons for this choice are motivated below (see
\S~2); in general, we find that the ray-shooting + PS approach constitutes a
flexible and accurate enough tool for our purposes.
The PS approach has been used also in conjuction with analytical techniques
by several authors (Narayan \& White 1988, Kochanek 1995, Nakamura \& Suto 1997).

This approach is somewhat different with respect to others used in past 
literature to investigate analogous problems. 
Schneider \& Weiss (1988) have applied the ray-shooting scheme to study the
propagation of light in a clumpy universe and to derive information on
the influence of matter inhomogeneities on the apparent luminosity of
lensed sources. The main difference with that work, is that we allow the lens
mass to be distributed according to the prescriptions
of a given cosmological model, rather than assuming identical clumps of mass $M$.  
Essentially for the same reasons, our work differs also from the one presented by
Lee \& Paczynski (1990), who consider GL in a flat spacetime and approximate the
true three-dimensional distribution of matter with multiple identical screens. 
Jaroszy\'nski (1991) allows for density fluctuation to grow starting from the
particular initial power spectrum $\propto k^{-3}$ (which is at the limit
of the hierarchical structure formation model) and superimposing a galaxy
population whose mass distribution is inferred from the Schechter function
and a prescription for the $M/L$ ratio. The delicate assumption that galaxies
trace mass in the universe, as well as the use of the Schechter distribution 
at high redshift, appears difficult to be justified, however.

Massive computational methods coupling the full power of N-body simulations
and ray-shooting start to become available in the literature (Wambsganss \etal
1996; Premadi \etal 1997), although with some unavoidable limitations which 
make them suitable for more focussed studies. 
Wambsganss \etal (1996) examined the evolution of various GL statistical 
indicators  (\ie multiple image frequency, lens redshift distribution,
magnification maps, caustic and critical line geometry) in a CDM
model up to $z=3$. In the second work, gravitational shear and magnification 
of a single light beam produced by lenses in the redshift range $z=0-5$ for 
three cosmological models are calculated. 
Our work is intended to mainly explore the magnification effects produced
by the cosmological matter distribution deduced from three different
hierarchical models (SCDM, LCDM, CHDM whose parameters are given in \S~2)
on very high redshift sources. Particular emphasis is devoted to the 
perspectives offered by GL for the observational detection of SNe in
protogalactic objects. We also explain how such observations could help
constrain the validity of the proposed classes of cosmological models. 

The plan of the paper is as follows. In the next Section we define
the cosmological models under study and derive their predicted lens
mass distribution; in \S~3 we introduce and discuss the basic concepts of 
thick gravitational lenses. Sec. 4 is devoted to the description of
the numerical scheme used for the GL simulations whose results are
presented in \S~5. Sec. 6 contains the observational implications of
our results for future detection of high-$z$ SNe with NGST; finally, the
discussion and summary in \S~7 conclude the paper.

\section{Cosmological Lens Distribution}

As stated above, our main aim is to determine the GL 
effects due to the intervening cosmological mass
distribution on distant sources. Currently, the most attractive
models for structure formation in the universe, assume that 
perturbations grow from an initial gaussian field of fluctuations
with a scale-free power spectrum. These perturbations become nonlinear
at redshift $\approx 50$ and form bound objects, which subsequently
merge to form larger and larger structures in a hierarchical fashion. 
In spite of the relatively widespread agreement on this general
scenario, a detailed comparison of the predictions of such type 
of models with observational data does not allow yet to discriminate
among the various hierarchical models so far proposed and to uniquely
fix their free parameters.  

Given these uncertainties, and with the hope to constrain at least 
some of the properties of the proposed models, we will consider
and compare three different cosmological models with the same total density
parameter $\Omega=\Omega_M + \Omega_\Lambda + \Omega_{\nu} =1$, contributed
by cold dark matter + baryons, cosmological constant $\Lambda$, and hot
dark matter, respectively. In particular, 
{\it (i)} Standard Cold Dark
Matter (SCDM), has $\Omega_M=1, \Omega_\Lambda=\Omega_{\nu}=0$;
{\it (ii)} Lambda Cold Dark Matter (LCDM) has  $\Omega_M=0.4,
\Omega_\Lambda=0.6, \Omega_{\nu}=0$; finally, {\it (iii)} Cold+Hot Dark
Matter (CHDM) parameters  are $\Omega_M=0.7, \Omega_\Lambda=0, \Omega_{\nu}=0.3$.

In order to specify the spectrum completely we have to fix both
the present value of the Hubble constant  $H_0= 100 h$~km~s$^{-1}$~Mpc$^{-1}$,
and the normalization of the spectrum, often parameterized by its
gaussian variance (see eq. \ref{sigmam} below) 
$\sigma_8=\sigma(8h^{-1}$~Mpc). The value of $h$ 
remains uncertain by about a factor of two: $0.4 \le h \le 1$.
Standard relative distance methods using Type Ia SNe 
(Garnavich \etal 1997, Perlmutter \etal 1998,  Kim \etal 1997) 
tend to favor higher values of $h\approx 0.6-0.8$, whereas methods
based on fundamental physics methods, including GL, suggest lower
values $h\approx 0.4-0.6$ (Grogin \& Narayan 1996; Kneib 1998 and
references therein).  
For this reason, we adopt  the intermediate value $h=0.65$ throughout
the paper; fixing $h$ has the advantage that, when comparing the 
predictions of the various models, one can isolate differences that depend
only on their intrisic properties listed above. 
The second point concerns the normalization of the power spectrum.
The actual form of the power spectrum can be obtained by numerically integrating
the relevant numerical equations; we use the analytical fit to such results 
provided by Efstathiou \etal (1992): 
\begin{equation}
\vert \delta_{k} \vert ^{2}= \frac{\displaystyle A k^{n}}{\displaystyle
\left\{1+ \left[Bk + (Ck)^{3/2} + (Dk)^{2} \right]^{\nu}
\right\}^{2/ \nu}},
\label{powerspectrum}
\end{equation}
where $\nu$=1.13; $B$=6.4 $h\Gamma$ Mpc; $C$=3.0 $h\Gamma$ Mpc; $D$=1.7
$h\Gamma$ Mpc, $n$=1; $\Gamma$ is the shape parameter and it is equal to
$\Gamma=h$ for SCDM, $\Gamma=h \Omega_M$ for LCDM, $\Gamma=0.2(0.3/\Omega_\nu)^{1/2}$ 
for CHDM. The coefficient $A$, or equivalently the value of $\sigma_8$, 
must be determined by choosing an appropriate normalization. We fix $\sigma_8$
such that the number of clusters present at $z=0$, about one per $(100 h^{-1}
Mpc)^3$, is correctly reproduced. This implies $\sigma_8 =0.64, 1.07, 0.68$, 
for SCDM, LCDM, CHDM, respectively.

\begin{figure}
\centerline{\psfig{figure=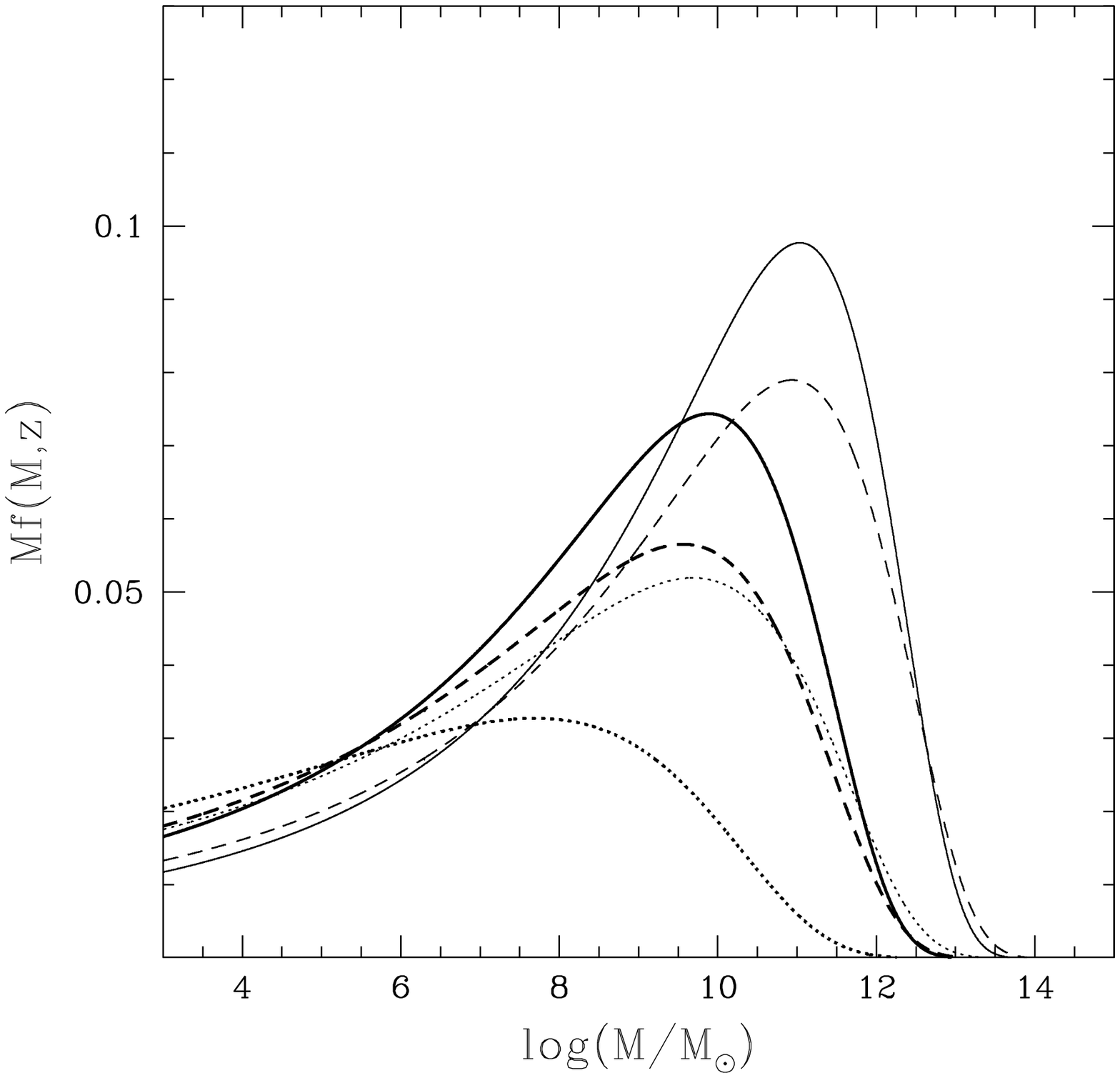}}
\caption{\label{fig1} Fraction of collapsed objects per unit logarithmic mass for
various cosmological models with parameters as given in the text: SCDM (solid 
line), LCDM (dashed) and CHDM (dotted) and for two different redshifts: $z=3$ (light
curves) and $z=5$ (thick).} 
\end{figure}

Once the spectrum has been fully specified, the structure formation and
evolution can be determined, for example, via N-body numerical simulations.
Here we take a complementary approach that makes use of the Press-Schechter
(PS) formalism (Press \& Schechter 1974; Bardeen \etal 1986; Peacock \& Heavens
1990; Bond \etal 1991) in its standard form; 
we will briefly discuss later in 
this Section  the main advantages and limitations of such approach.
Given a power spectrum $\vert \delta_{k} \vert ^{2}$ (in our case eq. 
\ref{powerspectrum})
one can write the gaussian variance of the fluctuations on the mass scale $M$:
\begin{equation}
\sigma_{M}^{2}=\int
\frac{d^{3}k}{(2\pi)^{3}}W^{2}(k,R)|\delta_{k}|^{2},
\label{sigmam}
\end{equation}
where $M=(4/3)\pi \rho R^3$, $\rho$ is the matter density, and 
\begin{equation}
W={3\over (kR)^3} \left[ \sin (kR)-(kR)\cos (kR) \right];
\label{wf}
\end{equation}
is a top-hat filter function.  From the results 
of the nonlinear theory of gravitational collapse,
stating that a spherical perturbation with overdensity $\delta_c = \delta \rho/
\rho > 1.69$ with respect to the background matter collapses to form a
bound object, through the PS formalism we can derive the normalized fraction
of collapsed objects per unit mass at a given redshift:
\begin{equation}
f(M ,z)=\sqrt{\frac{2}{\pi}}
\frac{\delta_{c}(1+z)}{\sigma_{M}^{2}}
e^{-\delta_{c}^{2}(1+z)^2/2\sigma_{M}^{2}}
\left( -\frac{d\sigma_{M}}{dM} \right).
\label{fmz}
\end{equation}
A useful quantity is the cumulative mass function  
\begin{equation}
F(M,z)=\int^{M}_{0} f(M ,z)dM = {\it erf}
\left( \frac{\delta_{c}(1+z)}{\sqrt{2}\sigma_{M}}
\right).
\label{fcmz}
\end{equation}
In Fig. \ref{fig1}, as an example, we show the function $ f(M ,z) M$, which
gives the fraction of collapsed objects per unit logarithmic mass for
the three cosmological models and two different redshifts.

The PS formalism has been shown to be in surprisingly good   
agreement with the results from N-body numerical simulations (Brainerd \& 
Villumsen 1992, Gelb \& Bertschinger 1994, Lacey \& Cole 1994, Klypin \etal 1995).
Nevertheless, it suffers from several limitations (for a discussion see, for
example, Yano \etal 1996; Porciani \etal 1996),
and that the complexity of structures in the universe cannot be
described in full detail. There are at least two pieces of information
that are not present in the PS recipe: (i) the internal structure, \ie
the density profile, of the collapsed objects, and (ii) their spatial distribution
at a given redshift. In general, these aspects are relevant in GL experiments
and for this reason, as we will discuss in the next Section, one has to  
make some additional hypothesis. Since we approach the study of GL effects
using statistical methods, the bias introduced by those assumptions needs to be 
evaluated carefully; we leave this to the final Section.
However, obtaining the mass distribution as a function of redshift with PS,
not mentioning its widespread use in literature and in GL studies that
allows for straight
comparisons and checks, has the advantage that it is computationally convenient
and therefore the parameter space exploration and the optimization of the GL
numerical scheme (see \S 4) can be performed with relatively unexpensive
procedures. For these reasons we feel that is worthwhile to attack the 
problem with these approximate techniques that can represent a useful complement
to more massive future computations.

\section{Thick Gravitational Lenses}

In this paper, we are interested in determining the GL magnification
properties of high redshift sources (typically SNe) due to the intervening
cosmological matter distribution. Since this matter distribution, as obtained
from N-body simulations and/or PS formalism, covers a large range of redshifts,
the standard assumption that the lensing mass is concentrated on a
single plane perpendicular to the propagation of the reference light ray (\ie the
optical axis) has to be modified
to include such spread. This can be achieved by splitting the mass distribution
on a series of planes, and furthermore postulating that (i) deflections on the
various planes are independent and (ii) the trajectory of a light ray is modified
only at planes.
These deflections can be calculated through the deflection potential
$\Psi_z(\xi_z)$, where $z$ refers to the redshift of the plane and $\xi_z$
is the impact parameter on that plane; the propagation between planes can be
generally described by the Dyer-Roeder (1972) equation, also discussed by
Futamase \& Sasaki (1989), Babul \& Lee (1991), and Nakamura (1997).
In brief, the main requirement when using Dyer-Roeder distances
is that the redshift intervals are $\simlt 1$, in order to fulfill the
underlying hypotheses of the theory, \ie a light ray bundle is not affected
by nonlocal gravitational effects (shear) and significative local deflections.
The validity of the above scheme, which pundits call a "Thick Gravitational 
Lens",  is discussed in detail by Seitz \etal (1994),
Seitz \& Schneider (1994), but also in  Schneider \etal (1992) and Schneider
(1997).
In the following we describe how we model the lens systems corresponding to
the various cosmological models using the above approximation. 

\begin{figure}
\centerline{\psfig{figure=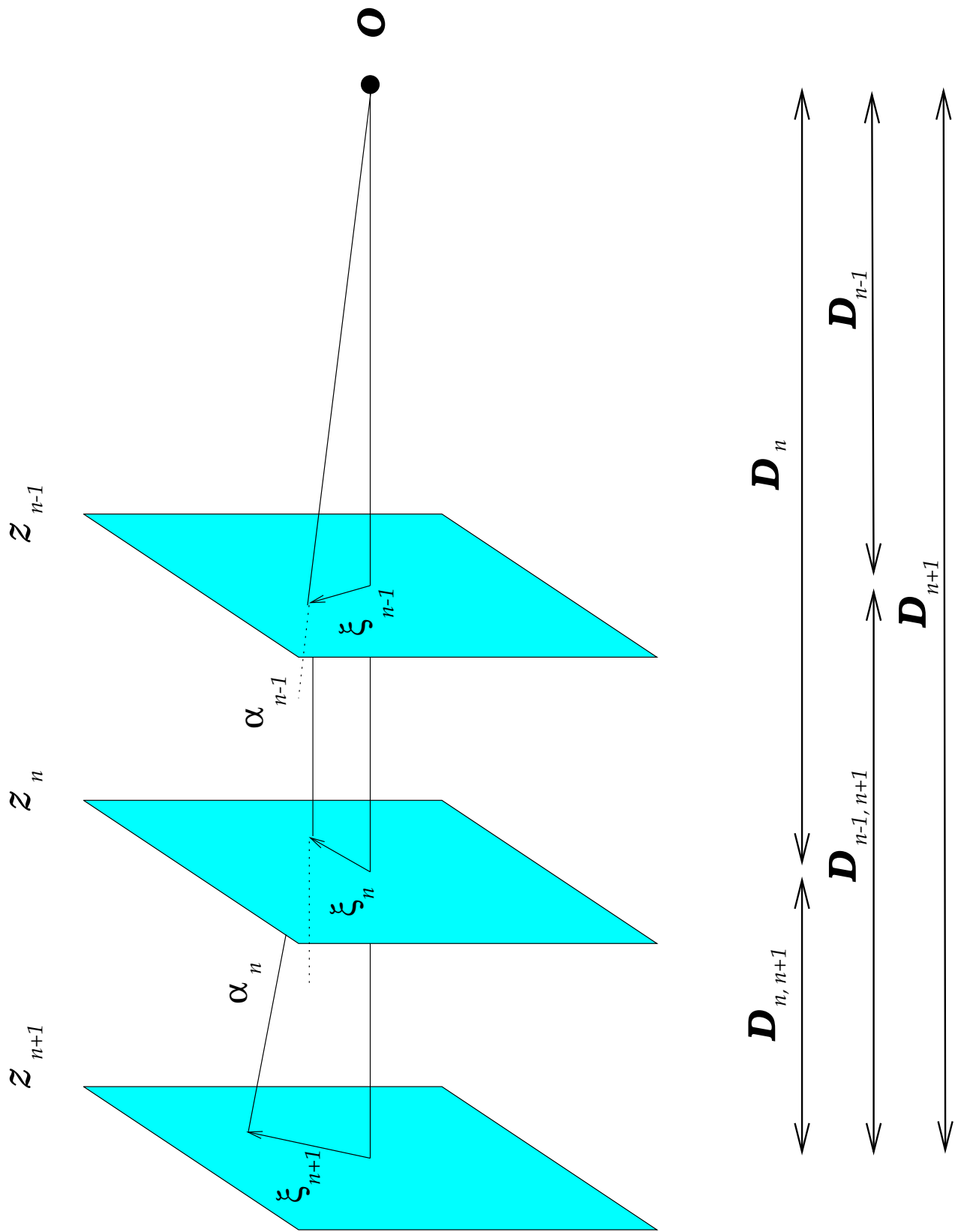}}
\caption{{\footnotesize\label{fig2} Schematic view of the Thick Gravitational Lens scheme
with symbols defined in the text.}}
\end{figure}

Let us suppose to split the mass on $N$ planes, at redshift $z_1, z_2, ..,z_n,..,
z_N$, equally spaced with redshift interval $\Delta z < 1$ both for the above reasons
related to the use of Dyer-Roeder equation, and to correctly sample the cosmic mass
distribution eq. \ref{fmz}. A sketch of such configuration
is given in Fig. \ref{fig2}.
Then the impact parameter on the plane $n+1$ is given by  
\begin{equation}
\xi_{n+1}=-\frac{(1+z_{n-1})D_{n,n+1}}{(1+z_{n})D_{n-1,n}}\xi_{n-1}
+\frac{D_{n-1,n+1}}{D_{n-1,n}}\xi_{n}
-D_{n,n+1}\nabla_\xi\Psi_{z_n}(\xi_{n}),
\label{xin1}
\end{equation}
where the deflection potential is 
\begin{equation}
\Psi_z(\xi_z)={4G\over c^2}\int d^{2}\xi^{\prime}\Sigma_z(\xi^{\prime})
\ln
\left( \frac{|\xi_z-\xi^{\prime}|}{\xi_{0}}
\right);
\end{equation}
$\Sigma_z$ is the mass surface density on the plane at redshift $z$ and
$\xi_0$ is an arbitrary length scale.
The angular distance $D_{ij}=D(z_i,z_j)$ is determined by solving the 
Dyer-Roeder equation in the form suggested by Linder (1988a, 1988b) 
\begin{equation}
\ddot{D}+\frac{3+q(z)}{1+z}\dot{D}+
\frac{3}{2}(1+z)^{-2}D
\sum_{s} (1+s ){\cal A}_{s}(z)
\Omega_{s}(z)=0.
\label{eqndr}
\end{equation}
Here the dot indicates redshift derivatives; $s$ is defined by the equation of
state $p= \Sigma_s s\rho_s$ involving the total pressure and the various 
density contributions, including the cosmological constant, in a Friedmann-like 
universe; for example, $s=-1$ for the cosmological constant, whereas $s=0$ for
a dust universe. Finally, a convenient expression for the generalized
deceleration parameter $q(z)$ is (Linder 1988a) 
\begin{equation}
q(z)=-\frac{a\ddot{a}}{\dot{a}^{2}}=
\frac{\sum_{s} \Omega_{s}(1+3s)
(1+z)^{1+3s}}{\sum_{s} \Omega_{s}
(1+z)^{1+3s}};
\label{qz}
\end{equation}
${\cal A}_s$ is the so-called Dyer-Roeder parameter that expresses the degree  
of homogeneity of the universe, \ie ${\cal A}_s=1$ corresponds to a perfectly 
homogeneous matter distribution, ${\cal A}_s=0$ corresponds to the case in which
all the matter is in bound systems. In general, eq. \ref{eqndr} can be solved
only numerically once the appropriate boundary conditions are specified;
in a few limiting cases, though, analytical solutions can be found. For the
case of interest here, in which the matter is supposed to be clumped in
a discrete number of collapsed objects 
projected onto the various planes, 
the relevant limiting case is ${\cal A}_s = 0$ for which the
following solution holds:
\begin{equation}
D(z_1,z_2)=(1+z_1)[D(z_2)-D(z_1)],
\label{inte2}
\end{equation}
where
\begin{equation}
D(z)={c\over H_0}\int_{1}^{1+z} \frac{dy}{y^{3}\sqrt{\sum_{s} \Omega_{s}
y^{1+3s}}}.
\label{inte3}
\end{equation}
This is the formula we use to calculate distances in eq.\ref{xin1}. 
We note that the relation between the Dyer-Roeder angular distance and the more
familiar luminosity distance, $D_L$ is always equal to $D= (1+z)^{-2} D_L$. 

The deflection angle caused by the n-th plane, 
$\alpha_{z_n}(\xi_{n})$ is defined as the gradient 
of the deflection potential entering eq.~\ref{xin1}. 
Throughout this paper we will assume for simplicity and for computational
economy that lenses are point-like masses, an approximation whose limitations
are discussed in \S~7. The mass distribution of this lens ensamble is
derived via the PS formalism (eq. \ref{fmz}). It is easy to verify that 
in this case
\begin{equation}
\alpha_{z_n}(\xi_{n}) = \nabla_\xi\Psi_{z_n}(\xi_n)= \frac{4GM_{*}}{c^{2}}
\sum_{j} \frac{M_{j}^{(n)}}{M_{*}}
\frac{\xi_{n}-\xi_{j}^{(n)}}
{|\xi_{n}-\xi_{j}^{(n)}|^{2}},
\label{alfa}
\end{equation}
where $M_*$ is a reference mass and $j$ runs on all the objects on the n-th plane; 
$\xi_j^{(n)}$ is the position of the j-th mass on the n-th plane, $M_j^{(n)}$.
In GL studies it is very common to use angular coordinates rather than
linear ones in order to define the deflection of light rays. We follow this
tradition by introducing the angular impact parameter $\theta_n=\xi_n/D_n$,
where $D_n$ is the distance from the observer to the $n$-th plane. Eq.~\ref{xin1} 
then becomes (also using eq.~\ref{alfa}) 
\begin{equation}
\theta_{n+1}=\frac{I_{n-1}^{n+1}}{I_{n-1}^{n}}
\theta_{n}-\frac{I_{n}^{n+1}}{I_{n-1}^{n}}\theta_{n-1}
-(1+z_{n})I_{n}^{n+1}\frac{4GM_{*}H_{0}}{c^{3}}
\sum_{j} \frac{M_{j}^{(n)}}{M_{*}}
\frac{\theta_{n}-\theta_{j}^{(n)}}
{|\theta_{n}-\theta_{j}^{(n)}|^{2}},
\label{thn1}
\end{equation}
where the auxiliary variable $I_{i}^{j}=(c/H_0)(D_i^{-1} - D_j^{-1})$ 
has been used.  From eq. \ref{thn1} one can appreciate the existence 
of a characteristic
redshift-independent value of $\theta$ which will be used as the natural unit
angle for the numerical simulations
\begin{equation}
\theta_{u}=\sqrt{\frac{4GM_{*}H_{0}}{c^{3}}}
\simeq 2\times 10^{-6}\sqrt{\frac{M_{*}}{M_{\odot}}}{\rm arcsec}.
\label{thu}
\end{equation}
Because of image splitting and distortion, GL produces a
magnification of the flux from a given source. This can be
understood by recalling that the
angular size of the source is increased and, since the surface
brightness is conserved, the net flux received is olso increased. 
In principle, we can calculate the total magnification $\mu$ (\ie including all 
the possible multiple images) of a source located, for example, 
on the plane $\theta_N$ iteratively by using eq.~\ref{thn1}: 
\begin{equation}
\mu^{-1}(\theta_1) = \det \left| \left| {\partial \theta_N\over \partial \theta_1}
\right| \right|
\label{mu}
\end{equation}
and summing over all multiple images. Regions in the source plane for 
which $\mu \to \infty$ are 
known as caustics. In practice, this method can be applied 
only to a very limited number of cases in which the lens equation can be
inverted, that is $\theta_1$ can be expressed as a function of $\theta_N$.
In the next Section we will describe how $\mu$ is calculated in practice
within the numerical scheme used to solve the problem.

\section{Numerical Simulations: Method and Description}

The problem in which we are interested here, \ie the magnification of distant
sources by the intervening cosmological distribution, and outlined in the 
previous Sections, cannot be solved by any analytical technique. Thus,
we have to resort to numerical simulations; in particular we adopt 
the so-called {\it ray-shooting} method. A full description of this scheme
can be found, for example, in Wambsganss (1990) (see also Schneider \& Weiss
1988); in the following we outline only
its main features. 

Let us consider a given cosmic solid angle, $\omega$, and study 
the propagation of light rays
through it; the main idea of ray-shooting is to follow such propagation
in a direction that is opposite to the physical one, namely from
the observer to the source. This propagation from a plane to the next
one is mapped by eq. \ref{thn1} above for each ray. Thus, illuminating 
all the images produced by the mass distribution contained in $\omega$, 
we can obtain an accurate description of the spatial pattern of the caustics
and calculate the source magnification.

We now determine the cosmological mass distribution to be projected
onto the various planes. We start by calculating the total mass contained
in the cone defined by the solid angle $\omega$. 
The comoving volume of such cone between
the two redshifts $z_1$ and $z_2$ is given by (Linder, 1988a; Carrol \etal
1992)
\begin{equation}
V(\omega,z_{1},z_{2})=
\omega (\frac{c}{H_{0}})^{3} \int_{z_{1}}^{z_{2}}
dz\frac{H_{0}}{H(z)}
\left[ \int_{1}^{1+z}
\frac {dy}{y\sqrt{\sum_
{s} \Omega_{s}y^{1+3s}}}
\right] ^{2},
\label{volume}
\end{equation}
where $H(z)$ is the Hubble constant at redshift $z$. The corresponding 
total mass is then 
simply ${\cal M}_{1,2}= (\Omega_M+\Omega_\nu)\rho_c V(\omega,z_{1},z_{2})$, 
where the critical 
density $\rho_c= (3 H_0^2/8 \pi G)$.
We then subdivide this mass among various point lenses 
distributed according to the prescribed mass distribution function  
obtained from PS (eq. \ref{fcmz}) and appropriate for the cosmological 
model (SCDM, LCDM, CHDM) under study. Specifically, 
we extract via a Monte Carlo procedure values of the masses from such
distribution, until their sum exceeds the value of ${\cal M}_{i,j}$ 
calculated for the two redshifts bracketing the redshift of the relevant plane.
This coupled PS+Monte Carlo procedure, being very flexible and handy, can be
used to extract from any hierarchical cosmological model the information
required for the setup of a  GL numerical experiment. 
Since the PS formalism does not provide us with any information about the
spatial distribution of the collapsed structures (we neglect in 
this paper any GL effect related to the possible different distribution of baryons),
it is necessary to make an {\it ad hoc} hypothesis concerning it.
The simplest assumption is that the lenses are spatially uncorrelated and randomly
distributed on the planes. Of course, this neglects the fact that in reality a
certain, albeit not completely understood, degree of clustering could be
present even at very high redshift. 
The limitation of this and the point lens
hypotheses, that are the main ones in this paper, will be further
discussed in \S 7.   

With this background, we are now ready to describe the main features of
the numerical computations and of the determination of the magnification 
probability of a lensed source.   
On the first plane, we define an orthogonal grid made of $N_l \times N_l$
pixels of size $\Delta_l$ and we shoot a light ray at each cell center. 
We then apply the lens equation \ref{thn1} to the angular impact position 
$\theta_1$ of each ray to obtain all the subsequent positons $\theta_2, \theta_n,
\theta_N$ up to the source plane. There we collect the final impact position 
in another grid made of $N_s \times N_s$ cells of size $\Delta_s$. The choice 
of $\Delta_s$ must be a compromise to minimize the effects of two
statistical biases: on the one hand the number of rays collected in each cell has to be
large enough to allow for a good statistical significance of averages;
on the other hand, cells that are too large to prevent good spatial resolution
of the caustic patterns. For this reason we constrain $\Delta_s$ to be 
\begin{equation}
\Delta_{l} \ll \Delta_{s} \ll L_{s},
\label{conditio}
\end{equation}
where $L_s = N_s \Delta_s$. In general, $L_s < L_l = N_l \Delta_l$ in order to
avoid well-known border effects, namely the flux attenuation in cells close 
to the grid borders due to the neglect of masses external to the cone. 
The final product of a numerical run is a $N_s \times N_s$ square matrix;  
each element of the matrix represents the number of counts, $n_{ij}$, 
in the cell $(i,j)$. The magnification matrix can then be obtained 
by evaluating the change in the actual number of rays per cell in the two grids:
\begin{equation}
\mu_{ij}(z)=n_{ij}(z)\frac{\Delta_{l}^{2}}{\Delta_{s}^{2}}.
\end{equation}
Due to the statistical character of the lens distribution,
the information contained in $\mu_{ij}(z)$ has to be interpreted
statistically, as well. This can be done by introducing the magnification
probability function
\begin{equation}
p(\mu,z)=\frac{1}{N_{s}^{2}}\sum_{i,j=1}^{N_{s}}
{\cal H} \left[ \mu_{ij}(z)-\mu \right] ,
\label{pmu}
\end{equation}
where ${\cal H}=1$ if  $\mu_{ij}=\mu$, and ${\cal H}=0$ otherwise.
An additional useful quantity is the cumulative magnification probability
function 
\begin{equation}
P(>\mu,z)=\frac{1}{N_{s}^{2}}\sum_{i,j=1}^{N_{s}}
\theta_{H}(\mu_{ij}(z)-\mu),
\label{Pmu}
\end{equation}
where $\theta_{H}$ is the Heaviside function.
Clearly, since $\mu$ is calculated for each cell, this implicitly implies 
that the probability distribution is  appropriate to point sources.

 
We conclude this Section giving the numerical values of the above parameters
used in the simulations.
We use $N = 50$ planes, equally spaced by a redshift interval $\Delta z = 0.2$,
thus spanning the total redshift range $z=0-10$.
The number of rays followed in each simulation is $N_l^2=1.85\times 10^7$;
they are uniformly distributed over a solid angle $\omega = 2.8\times
10^{-6}$~sr, corresponding to a $420 \times 420$~arcsec field. We assume
$M_* = 10^{15} M_\odot$ (corresponding to the largest mass predicted by PS
in each run) and consequently $\theta_u = 42$~arcsec (see eq. \ref{thu}).
Then it is $L_s= 7\theta_u = 294$~arcsec and, as stated above, $L_l=
10\theta_u = 420$~arcsec; $\Delta_s= L_s/N_s = 0.96$~arcsec for $N_s=300$. 
Since we are shooting one ray per cell on the first plane, the minumum angular
resolution is $L_l/N_l = 0.01$~arcsec, a typical deflection caused by a
mass $M\approx 10^{10} M_\odot$. Mainly for this reason we have applied a 
lower mass cutoff to the mass distribution $M_c = 5\times 10^{11}
M_\odot$ which represents a good compromise when computational economy is
taken into account. Of course, the number of lenses varies with redshift; 
the largest value in a plane is $ \approx 600$ lenses with little difference 
among different cosmological models. 
Finally, we point out that, although
the solid angle $\omega = 2.8\times 10^{-6}$~sr considered in the numerical 
runs is relatively small, the statistical significance of the
results is guaranteed by the sufficiently large number of masses typically
contained on each plane, of order of several hundreds.

\section{Results}

Our results are essentially contained in the magnification maps obtained
from the numerical simulations. In the following, we first present and describe
them in general terms, underlying the differences among cosmological models; 
next, we analyze them in detail by means of the magnification distribution
function described above.

\subsection{Magnification Maps}

\begin{figure}
\centerline{\psfig{figure=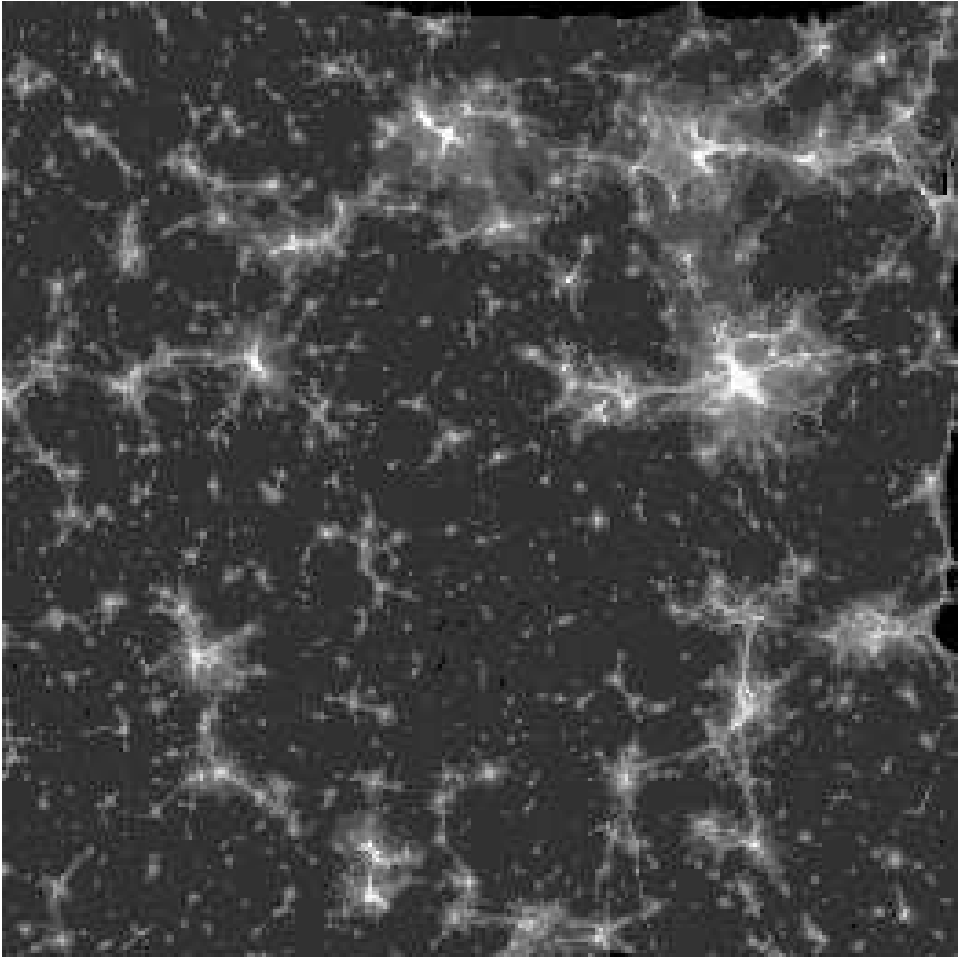}}
\caption{{\footnotesize\label{fig3} Magnification maps for the three cosmological 
models: (a) SCDM (b) LCDM (c) CHDM at redshift $z=6$
with symbols defined in the text.}}
\end{figure}

Fig. \ref{fig3} shows three magnification maps relative to the three
cosmological models under study. 
Typically, $N=50$ maps are obtained by gradually increasing the redshift of the lensed
point source mimicking a distant SN, possibly occurring in a Pop~III object.
Here we show as an example those corresponding to $z=6$ for each model, 
which nevertheless are representative of the general lensing features of the 
various cosmologies.  These maps give the magnification of the flux source as
a function of its putative position in the source plane.
Note that, since we have normalized all fluctuation spectra such to reproduce 
the observed number of clusters at $z=0$, the three models are very similar
at low redshift and start to differentate as the source is moved back in time.
Hence, the magnification patterns at sufficiently high $z$ truly evidentiate 
intrinsic differences in the mass distribution function predicted by the various 
models, allowing for a meaningful comparison.

Starting from the map corresponding to the SCDM model, we note that caustics
appear rather intense and concentrated in structures with sizes of the
order of $\approx 12$ cells, or $\approx 11$~arcsec. 
About 10 high magnification regions with $\mu \approx 50$ can be identified,  
which are produced by the most massive lenses encountered by the light rays 
along their cosmological path. The median magnification calculated on the entire 
map is $\bar \mu = 1.5$.
The LCDM magnification map in Fig.\ref{fig3}b, to a first sight looks similar to
the previous one, again showing the same filamentary structures. 
Intense caustics are less frequent and of smaller size; a very strong
magnification $\mu \approx 120$ is found at one location, but the median
magnification value is slightly lower than for SCDM, $\bar \mu = 1.47$.  
CHDM models, are instead peculiar. In fact, they produce a ``grainy"
pattern distribution, constituted by a large number of low $\mu$ caustics
of small angular size, of the order of a few grid points ($\sim$ few arcsec).
Caustics more intense than $\mu \approx 30$ are absent, and the median
magnification is $\bar \mu = 1.4$.  

These differences (or resemblances) among various models can be understood 
easily on a qualitative basis, noting that CHDM models predict 
less massive objects than SCDM or LCDM at any $z \simgt 0.5$.
This emerges in the different effects of GL which,
in the first case is dominated by the effects of masses typical of 
galaxies or small groups ($M\approx 10^{12} M_\odot$), whereas for 
cold dark matter only models is influenced by the presence of 
cluster-type lenses ($M\approx 10^{14} M_\odot$). The less sensible 
differences between SCDM and LCDM models arise mainly from the fact 
that for a given redshift the typical lens mass scale in the LCDM is 
about one order of magnitude smaller (see Fig.\ref{fig1}), whereas the
lens number density is roughly the same in the two models. 
This implies lower magnifications and caustic sizes, but apparently 
similar magnification patterns.

\subsection{Magnification Probability}

Fig. \ref{fig5} shows the cumulative distribution function $P(\mu, z)$
for the three models. Such distributions present a moderate degree of evolution
up to $z\approx 5$ (CHDM) and $z\approx 7$ (SCDM/LCDM). This behavior, as already 
stressed before, reflects the different trends of the mass function with
redshift. All models predict that statistically large magnifications, $\mu \simgt
20$ are achievable, with a probability of the order of a fraction of percent,
the SCDM model being the most efficient magnifier. Also the shape of the various
curves is rather similar. In order to appreciate better the differences among
the models and to quantify the magnification probabilities in more detail,
we introduce two parameters derived from the parent
distribution $P(\mu,z)$. These are: (i) $\mu_{10}(z)$, or the value of $\mu$
with a probability $P(\mu,z) = 10$\%, which gives an idea of the magnification
of the sources that are more likely to be detected, and (ii) $P(\mu > 10,z)$,
or the probability associated with high magnifications (we have set the 
threshold at $\mu=10$ to parallel the previous analysis in terms of the
two-point correlation function), which estimates how likely the detection of
highly magnified sources will be.
The evolution of these two parameters is illustrated in Figs.
\ref{fig6}-\ref{fig7}. From the behavior of $\mu_{10}(z)$, we see once again
that CHDM models produce (statistically) lower magnifications than the other 
two models. 
Nevertheless, all cosmologies predict that above $z\approx 4$ there is a 
10\% chance to get magnifications larger than 3, which, observationally,
corresponds to a magnitude gain of $\Delta m \approx 1.2$. Also, pushing
the realm of observations to very high $z$ potentially allow discrimination
of the models, or at least to discard some of them, provided that enough
lensed point sources can be detected to reproduce $P(\mu, z)$. These (and
other) observational implications are discussed in the next Section.
Similar considerations hold also for the evolution of $P(\mu > 10, z)$,
which shows a steep increase of the probability for all models, followed by
an almost constant plateau at about 1\% probability. Thus, in a field of
about 25 square arcminutes we expect that statistically, an area of at least 
$30'' \times 30''$ will produce magnifications larger than 10 times, corresponding
to a magnitude gain of $\Delta m = 2.5$.
Finally, it is interesting to note that at low redshift $z \simlt 3$ there is
a close resemblance in the magnification evolution of LCDM and CHDM models,
which differentiate only at earlier epochs. 


\section{Observational Implications for NGST}

According to hierarchical models of structure formation the first
luminous objects in the universe should have total masses
of order $\approx 10^6 M_\odot$ and form at redshift $z\approx 20-30$;
larger objects then form by merging of these building blocks.
However, a time gap between the first and subsequent generation 
of structures could be present, due to the possible feedbacks (H$_2$
photodissociation, reheating of the IGM due to stellar energy injection:
see Shapiro \etal 1994, Ferrara 1998).
Thus, for quite a long cosmic time interval these small objects could be 
the only visible trace of galaxy formation at the redshifts that we hope
to be able to unveil in the near future.
Given their small mass, these objects are likely to be faint: for a reasonable
mass-to-light ratio for young galaxies $\approx 0.1$, their luminosity is 
$\approx 2\times 10^{40}$~erg~s$^{-1}$. A Type II SN is typically
one hundred times brighter. Thus, for periods even longer than a year (taking
into account the time stretching $\propto (1+z)$ of the SN light curve)
the SN outshines its host galaxy. Since Type II SNe         originate
from massive stars, a necessary requirement is that the (unknown) IMF of
these objects is flat enough to extend into this regime. Clearly, this
condition is satisfied in the local universe, but it is not necessarily
so in the conditions prevailing when the universe was young. Nevertheless, it
is intriguing to speculate about the observational perspectives to detect
very high $z$ SNe, a possibility to which our hopes to
investigate directly the primeval star/galaxy formation are closely tied.

The advances in technology are making available a new generation of
instruments, some already at work and some in an advanced design phase, 
which will dramatically increase our observational capabilities.
As representative of such class, we will focus on a particular instrument,
namely the Next Generation Space Telescope (NGST). In the following we
will try to quantify the expectations for the detection of high $z$ SNe
and the role that the gravitational lensing can play in such a search.

We will assume that NGST (i) is optimized
to detect radiation in the wavelength range from $\lambda_{min}=1\mu$m to 
$\lambda_{max}=5\mu$m (\ie J-M bands), and (ii) can observe to a 
limiting flux of ${\cal F}_{NGST}=10$~nJy in $10^{2}$ s in that range,
which should allow for low-resolution spectroscopic follow-up.
We also assume that a Type II SN has a black-body spectrum (Kirshner 1990)
with temperature
$T_{SN}$ and we fix its luminosity $L_{SN}=3\times 10^{42}$ (Woosley \&
Weaver 1986; Patat \etal 1994).
This constant luminosity plateau lasts for about $\approx 80 (1+z)$~days, 
after which it fades away. 

Fig. \ref{fig8} shows the AB apparent magnitude of a SN as a function of
its explosion redshift in four wavelength bands (J, K, L, M) in the assumed NGST 
sensitivity range, and we compare it with the instrument flux limit. 
Note that we have taken into account absorption by the IGM 
at frequencies higher than the hydrogen Ly$\alpha$ line.
The plot also assumes that $T_{SN}=25000$~K, a value corresponding to the
temperature approximately appropriate to the first $15 (1+z)$~days after 
the explosion (Woosley \& Weaver 1986). After that, the temperature
drops to $T\simeq 6000$~K, where hydrogen recombines, for the entire duration
of the plateau (see below).  From an inspection of Fig. \ref{fig8} we deduce that,
without considering GL magnification effects, the J band offers the best
perspectives to observe such an object: at the assumed sensitivity limit, 
NGST should be able to reach $z \approx 4$. Including GL magnification 
enhances dramatically the observational capabilities. In fact, as one can
see from the solid line set of curves in Fig. \ref{fig8}, even allowing
for a magnification $\mu =3$ only, which in all models has a magnification
probability larger than 10\% at high $z$ (see Fig.~\ref{fig6}), this pushes
the maximum redshift at which SNe can be detected up to $z \approx 9$ 
in the K band. 

A different way to appreciate the GL effects is illustrated by Fig. \ref{fig9}.
There we allow for the SN temperature to vary in the large interval $T_{SN}
\approx 4000 - 10^5$~K and we ask which is the region of the  $(1+z)-T_{SN}$ 
parameter space in which NGST could possibly detect SNe in primordial objects.
This region is constrained by the requirement that the radiation from the
SN (or, indicatively, the maximum of its black-body spectrum) falls in between 
$\lambda_{min}$ and $\lambda_{max}$ with a flux larger than ${\cal F}_{NGST}$.
In the absence of magnification, the upper redshift boundary of the detection area 
(marked as "observable" in the Figure) is $4 < z< 8$ for $10^4~{\rm K}\simlt
T_{SN} \simlt 3\times 10^4~{\rm K}$ for SCDM and CHDM models (both having
$\Omega_\Lambda=0$) and $3 < z< 5.5$ for $8000~{\rm K}\simlt
T_{SN} \simlt 2.5\times 10^4~{\rm K}$ for LCDM. This difference is due to the
larger luminosity distance in LCDM models. We can now compare this result with the one
obtained considering the GL magnification of the SN, as calculated in the
previous Sections. As an indication, we use for $\mu$ the value of $\mu_{10}(z)$
shown in Fig. \ref{fig6}: this quantity should represent a good compromise
between the two, usually conflicting, requirements that a high magnification and high 
probability event occurs. In this case, the upper redshift boundaries (the three
uppermost declining curves in Fig. \ref{fig9}) of the
detection area are shifted towards higher redshift, in a way that depends 
on the cosmological models, which are now clearly differentiated because of 
the concomitant effect of their different $\mu_{10}(z)$ and luminosity distances. 
Also, the total area is generally increased, thus considerably enhancing the
detection chances. The maximum redshift at which a SN can hopefully be observed
is now above $z=10$, the limit of our GL simulations, for all models, and it
can be extrapolated to $z \approx 12$ for SCDM.
The most favorable SN temperature to maximize the depth of the search is $T_{SN}
\approx 14000$~K, but a large range, $10^4~{\rm K}\simlt
T_{SN} \simlt 3\times 10^4~{\rm K}$, allows to explore the universe illuminated
by SNe at redshifts larger than eight. This range is limited at high $T_{SN}$
by integalactic absorption of SN photons above the hydrogen Ly$\alpha$ 
frequency.

So far, we have emphasized how GL magnification can help us detecting
very high-$z$ SNe. However, magnification maps can also
contain precious information to discriminate among different cosmological 
models. There are at least two ways in which one can think of exploiting such
information.
In principle, the interpretation of SN number counts based on the 
results shown in Fig. \ref{fig9} could already provide a first insight
on the different properties of the models, since the 10\% magnification
probability contours are located at different positions. It follows that one 
would expect different decreasing trends in the number counts with redshift
above the $\mu=1$ line. However, this method crucially depends on the
star formation/SN production history in the universe; hence a firm 
interpretation of the results would require a good knowledge of such quantity. 

A different, more promising, strategy is instead the reconstruction of
the (cumulative) probability distribution, $P(\mu,z)$, shown in Fig. \ref{fig5}. 
Suppose that, by monitoring a given field (the planned NGST field of view 
has a size $240 \times 240$~arcsec, very similar to the $294 \times 294$~arcsec
maps shown in Fig. \ref{fig3}) at some appropriate time interval that depends on
redshift, a statistically significant number of SNe are discovered following 
their light curves. From the spectral energy distribution, say in the J K L M 
bands,  and assuming
a black body spectrum for the SN, one can determine the ratio $(1+z)/T$. 
If follow-up spectroscopy is available, the redshift (and, consequently, the
temperature) can be determined straightforwardly. However, even if this measure
is not possible, one can try a different strategy based on the fact that
the light curve is stretched by a factor $(1+z)$. It is well known that Type II
SNe fall essentially in two categories, characterized by the presence of a
luminosity plateau (SNeII-P) or by an almost linear luminosity decrease (SNeII-L).
The duration of the plateau in SNeII-P has a distribution which is 
peaked around the mean value $70$~days. Hence, lacking a 
better estimate, the redshift can be determined directly by properly sampling 
the light curve.  Obviously, the accuracy of this method cannot compete with 
a spectroscopic determination. Nevertheless, there is a fortunate coincidence 
inherent to the properties of the magnification pattern that makes this error
not so crucial when comparing different cosmological models. In fact, we
see that for redshift $z \simgt 5$ the $P(\mu,z)$ curves approach a "limiting"
curve (see Fig. \ref{fig7}), \ie there is no magnification evolution, appropriate 
for each model.
This is due to the fact that in hierarchical models objects present at such high
$z$ essentially do not contribute to further magnification because of their
low mass. It follows that, even if the redshift is not very precisely known
we can still reconstruct $P(\mu, z)$ to a very good accuracy by binning
together all the detected SNe with $z \simgt 5$. This is clearly an advantage 
of the magnification method with respect to the standard direct methods based 
on the luminosity distance for which a precise determination of the redshift is
crucial, as far as discriminating among cosmological models is concerned. 
The final step is represented by the determination of the absolute luminosity
of the plateau phase from which the magnification follows directly. 
Again we are facing the fact that Type II SNe are not
proper standard candles: althoug the plateau luminosity is observed to vary only 
by a factor $\approx 2$, there are exceptions to this rule (Patat \etal 1994). 
With this note of caution, we can have at least a first estimate of the
magnification.
 
We conclude this Section by mentioning that 
Type II SNe have also been used as distance indicators
via the so-called  Expanding Photosphere Method (Kirshner \& Kwan 1974, Montes \&
Wagoner 1995), which has been rather successful in distance determinations of 
closer SNe; this technique has recently been adapted for application to moderate
redshift SNe (Schmidt \etal 1994a). In short, the luminosity distance (and hence
constraints on cosmological models) can be
obtained from the formula
\begin{equation}
\label{EPM}
D_L(z) = \sqrt{{\zeta_\lambda^2 \pi R^2 B_\lambda(T)\over  \mu(z) f_\lambda}},
\end{equation}
where $T$ is the SN's observed color temperature, $\mu(z) f_\lambda$ is the observed
magnified flux density, $B_\lambda(T)$ is the Planck function and $\zeta_\lambda$
is a distance correction factor derived from models to account for
the dilution effects of scattering atmospheres; the photosperic radius, $R$, 
can be obtained from optically thin lines and considering free expansion of the
ejecta; for this reason, spectral data are necessary (Schmidt \etal 1994b). 
Thus, this method is suitable for redshift low enough that $\mu(z) \approx 1$;
for the redshifts of interest here, a model for the flux magnification has to be 
nevertheless assumed. 

\section{Discussion and Summary}

In this paper we have explicitly calculated the flux magnification
of distant sources  due to gravitational lensing effects of the intervening
matter distribution in the universe for different cosmological models.
We have particularly concentrated on Type II SNe, since at very high $z$
they are likely to offer the best detection perspectives, thus giving us
a clue on the star/galaxy formation history in the young universe.

We now discuss briefly the approximations and assumptions made.
We have assumed that the lenses are point masses. Although this 
approximation is commonly used in GL problems and numerical simulations
(Schneider \& Weiss 1988) due essentially to its simplicity, it remains an idealization
of the true lens matter distribution. In reality, even maintaining the less 
restrictive spherical symmetry hypothesis, lenses are more likely to be extended 
objects with a given radial density profile. Often, this is taken to be an 
isothermal profile, since this distribution tends to reproduce reasonably well the
properties of galaxy clusters (Narayan \& Bartelmann 1997) or leads to a
good agreement between COBE data and CDM cosmological models  (Kochanek 1995).
It is well known (Vietri 1985) that an isothermal sphere has generally
a lower magnification power at high magnifications than a point-like lens of the
same total mass.
Thus, the magnifications we find could be overestimated by some factor, which
depends on the redshift of the source. The magnitude of such effect is hard
to estimate since we are considering a thick gravitational lens and it is not
obvious how the magnification is modified with respect to the
single object case, for which this factor is typically $< 10$, depending
also on the compactness of the lens mass distribution.
It has to be pointed out that, in any case, a considerable debate is 
present in the literature on the detailed shape of dark matter halos.
CDM models predict that halo
density profiles are essentially self-similar with a weak mass
dependence (Lacey \& Cole 1994; Navarro \etal 1997); specifically the
dark matter density should decrease with radius as $\rho_h^{CDM}(r) \propto
r^{-1}$ at small radii,  down to the resolution of the simulation.
Recent observational work, however, appears to disagree with this
prediction.  Salucci \& Persic (1997) have suggested 
that dark matter halos have a core, \ie a central region of almost
constant density, whose size increases with galaxy luminosity both in
absolute units and as a fraction of the optical radius. 

The second approximation is the neglect of a possible clustering
of the lenses. Since this information cannot be derived directly
from the PS formalism used here, an additional hypothesis is necessary
to assign the spatial distribution of the lenses on the various planes;
as a first step we have adopted a random distribution. 
Again, reality might be different. Recent works (Hamilton \etal 1991, 
Matarrese \etal 1997) have been rather successful in reproducing 
observational data by detailed modelling of the clustering evolution
of various populations of objects (QSOs and galaxies). These results 
are dependent on the bias evolution which is still only partly
understood (Mo \& White 1996, Kauffmann \etal 1997, Catelan \etal 1997).
A detail study of the effects of clustering of galaxies on the 
magnification distribution function has been carried out by Jaroszy\'nski 
(1991). From Fig. 4 of that paper, and from the related discussion,
it is seen that the differences with respect to a randomly distributed
galaxy/lens population are negligible up to values of $\mu \approx 30$.
In the high magnification limit, a random distribution leads to a 
relatively lower probability: for example, for $\mu \approx 100$,
$P(\mu)$ is about 3 times smaller in that case. 
In a subsequent related and more refined work (particularly with a 
better statistics), Jaroszy\'nski (1992) instead found that the
discrepancy between random and correlated lens models are even 
weaker, and their associated magnification probability distributions 
would be undistinguishable for flat cosmological models.
The likelihood of observing distant supernovae depends only on the 
magnification probability. As our results have been calculated assuming
a magnification with a probability of 10\%, typically corresponding to
$\mu = 2-4$ (see Fig. \ref{fig6}), it follows from the above discussion,
that they should not be sensitive to the particular lens spatial 
distribution adopted. Instead, $P(\mu)$ depends more strongly upon
the lens mass distribution, as it is clear from the differences
among the three cosmological models studied here.
It is interesting to note that in any case as lens correlations tend 
to enhance high magnification probabilities (Jaroszy\'nski
1992), it will counterbalace to a certain extent the effect of  the point-like 
lens approximation discussed above. 

Finally, our description contains two additional hypotheses that 
should have only a minor effect on the results: 
(i) the baryonic mass fraction is the same in all lensing objects
(ii) we neglect the possible occurrence of microlensing due to stellar 
mass objects in intervening galaxies.

To conclude the discussion, we comment on our choice of Type II SNe as
the primary focus of our efforts. As we have already stated above these
events could represent the most distant observable ones in the early universe.
This statement is based on the recognition that a SN is expected to be
at least 100 times more luminous than their host Pop III object.
This is even more true for Type Ia SNe, that are on average brighter by
about 1.5 mag; moreover, Type Ia SNe are known to be very good standard candles
and, for this reason, they are widely used to determine the geometry of
the universe (Garnavich 1997; Perlmutter 1998). Hence, they could be 
excellent candidates for the present purposes. Unfortunately, it is reasonable
to expect that this events at $z\approx 10$ are very rare, since they arise
from the explosion of C-O white dwarfs triggered by accretion from a companion;
this requires evolutionary timescales comparable with the Hubble time at that 
redshift. A different possibility would be represented by Type Ib SNe, which
originate from short-lived progenitors: however, they share problems similar to
Type II SNe, \ie they are poorer standard candles and are fainter than Type
Ia SNe. In any case, our results are valid for all three types of SNe since
they give the relative flux enhancement due to GL magnification.

The summary of the main results of this paper, under the assumptions made
and discussed above, is the following:

$\bullet$ The magnification patterns are different for the three cosmological 
models considered: for a given $z$, caustics are more intense and concentrated 
in SCDM models and their pattern becomes more ``grainy'' for LCDM and,
particularly, for CHDM. Median magnifications are $\simeq 1.5$ for all models.

$\bullet$ The magnification probability function presents a moderate degree of evolution
up to $z\approx 5$ (CHDM) and $z\approx 7$ (SCDM/LCDM). All models predict that
statistically large magnifications, $\mu \simgt
20$ are achievable, with a probability of the order of a fraction of percent,
the SCDM model being the most efficient magnifier.
All cosmologies predict that above $z\approx 4$ there is a
10\% chance to get magnifications larger than 3.

$\bullet$ We have explored the observational implications of the above
results taking NGST, with the planned instrumental characteristics, as 
a reference instrument. We find that NGST should be able to detect and 
confirm spectroscopically Type II SNe up to 
a redshift of $z \approx 4$ in the J band (for $T_{SN}=25000$~K); 
this limit could be increased
up to $z\approx 9$ in the K band, allowing for a relatively moderate magnification. 
The detection area in the redshift-SN temperature plane depends on cosmological
models. When magnification is taken into account, and for the most
suitable values of $T_{SN} \approx 14000$~K, the maximum possible 
detection limit is pushed well
above $z=10$. Possibly promising strategies to discriminate among cosmological models
using their GL magnification predictions and very high-$z$ SNe are sketched. 

\vskip 2truecm
We acknowledge C. Kochanek, L. Moscardini, C. Porciani, P. Schneider,
M. Vietri for helpful comments.


\newpage
\footnotesize

\begin{figure}
\centerline{\psfig{figure=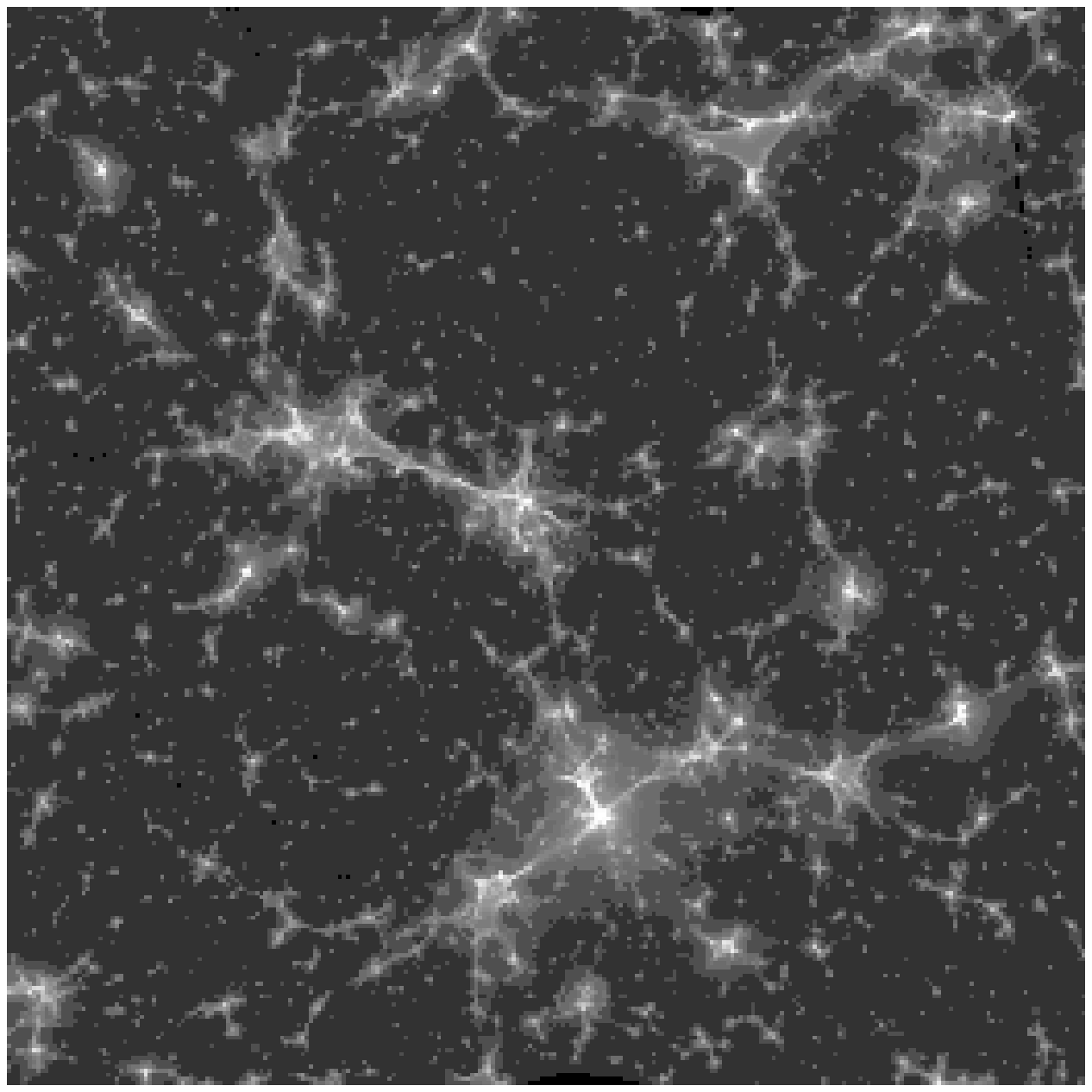}}
Fig. 3b -
\end{figure}

\begin{figure}
\centerline{\psfig{figure=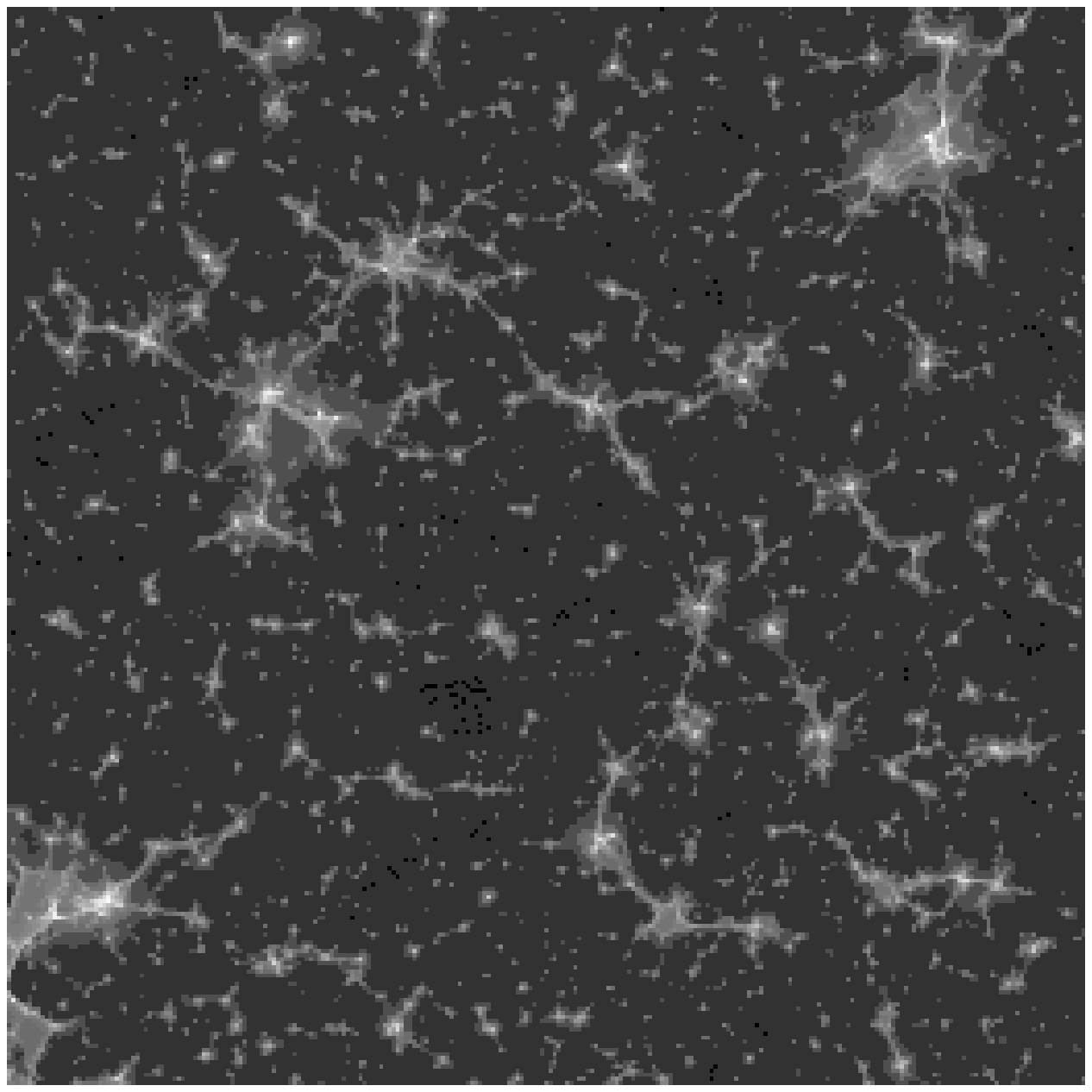}}
Fig. 3c -
\end{figure}
%
%
%
\begin{figure}
\centerline{\psfig{figure=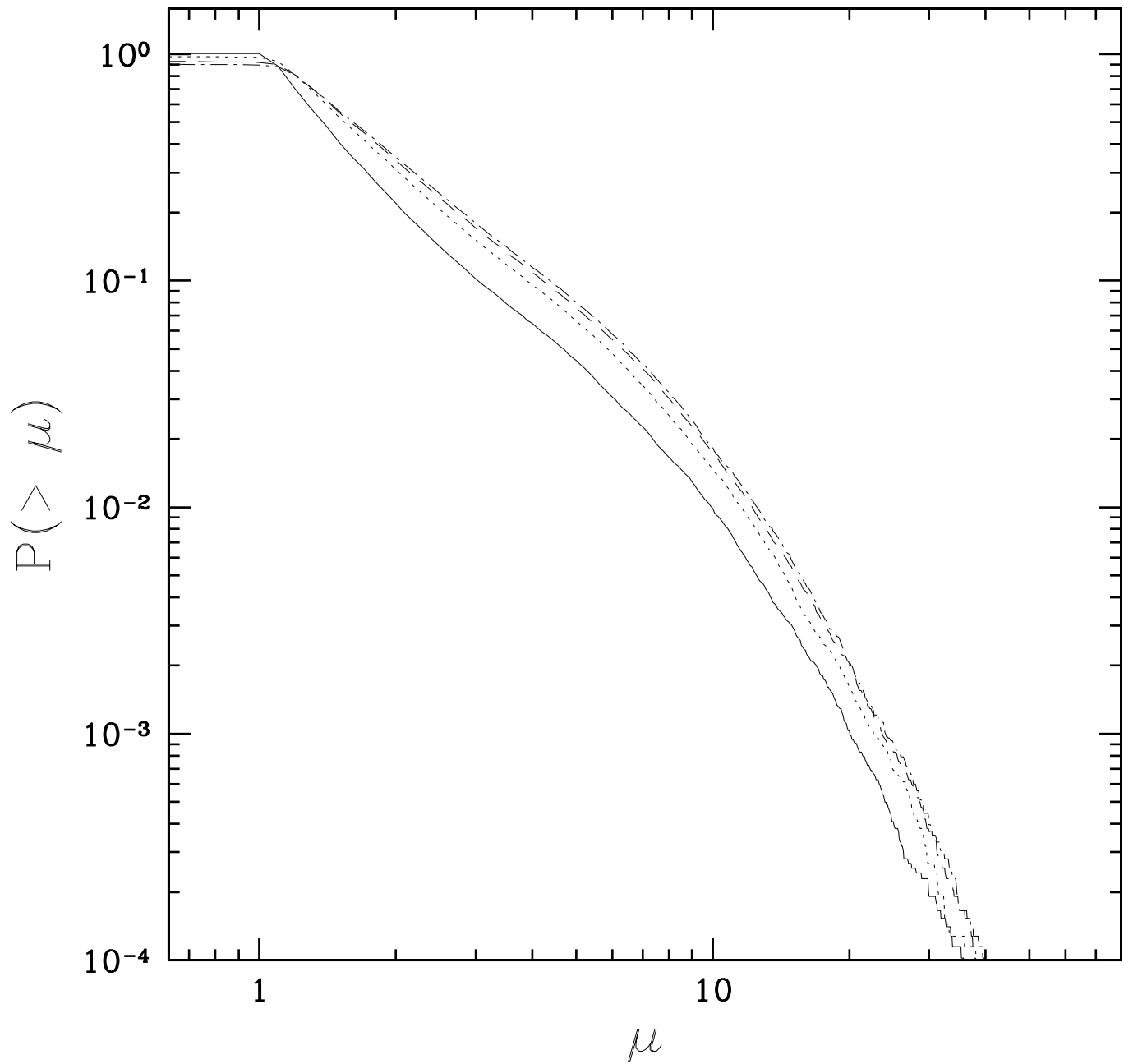}}
\caption{{\footnotesize\label{fig5} Cumulative magnification probability for four different 
redshifts $z=3$ (solid line), $z=5$ (dotted),  $z=7$ (dashed), and
$z=9$ (dot-dashed) for (a) SCDM, (b) LCDM, (c) CHDM models, respectively.}}
\end{figure}

\begin{figure}
\centerline{\psfig{figure=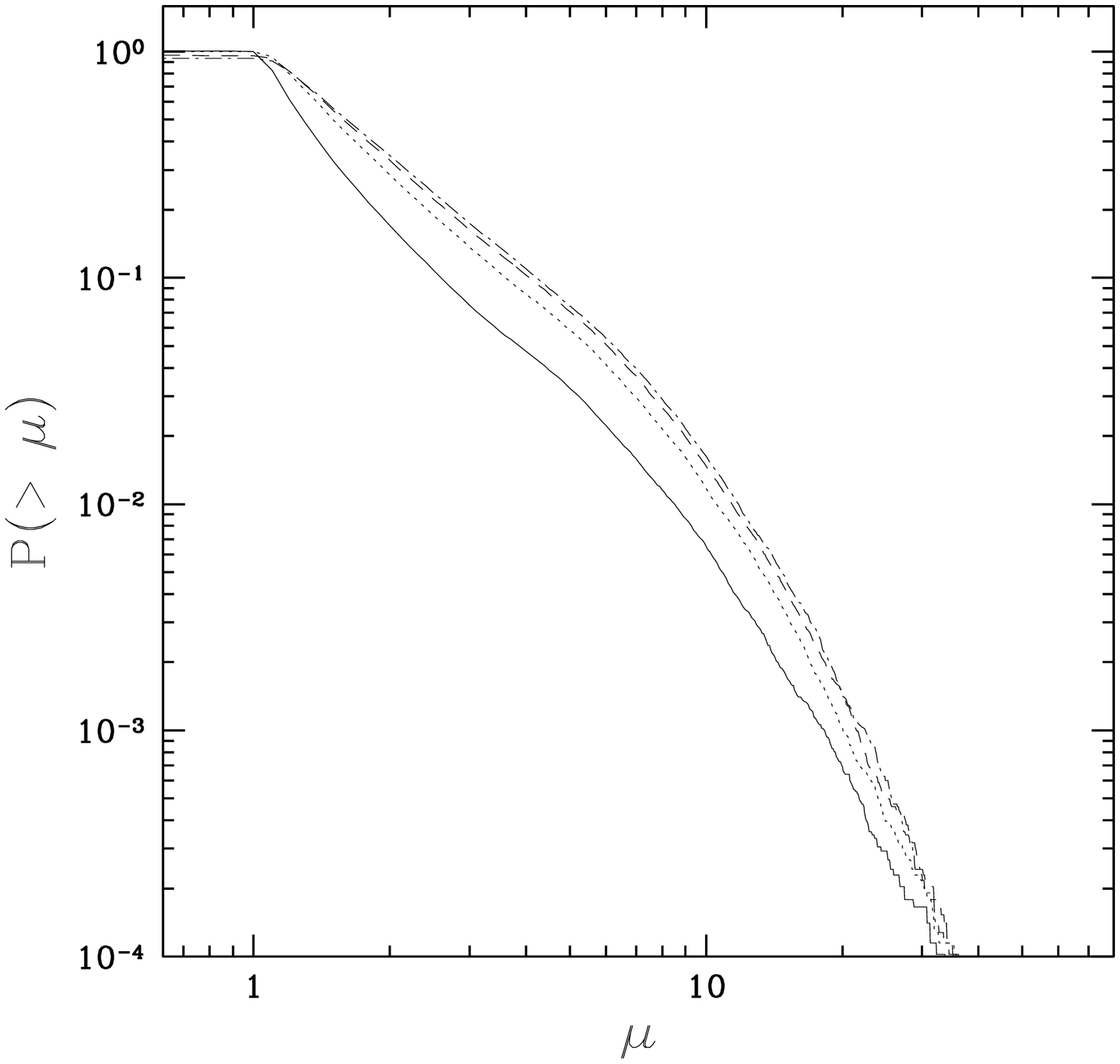}}
Fig. 4b -
\end{figure}

\begin{figure}
\centerline{\psfig{figure=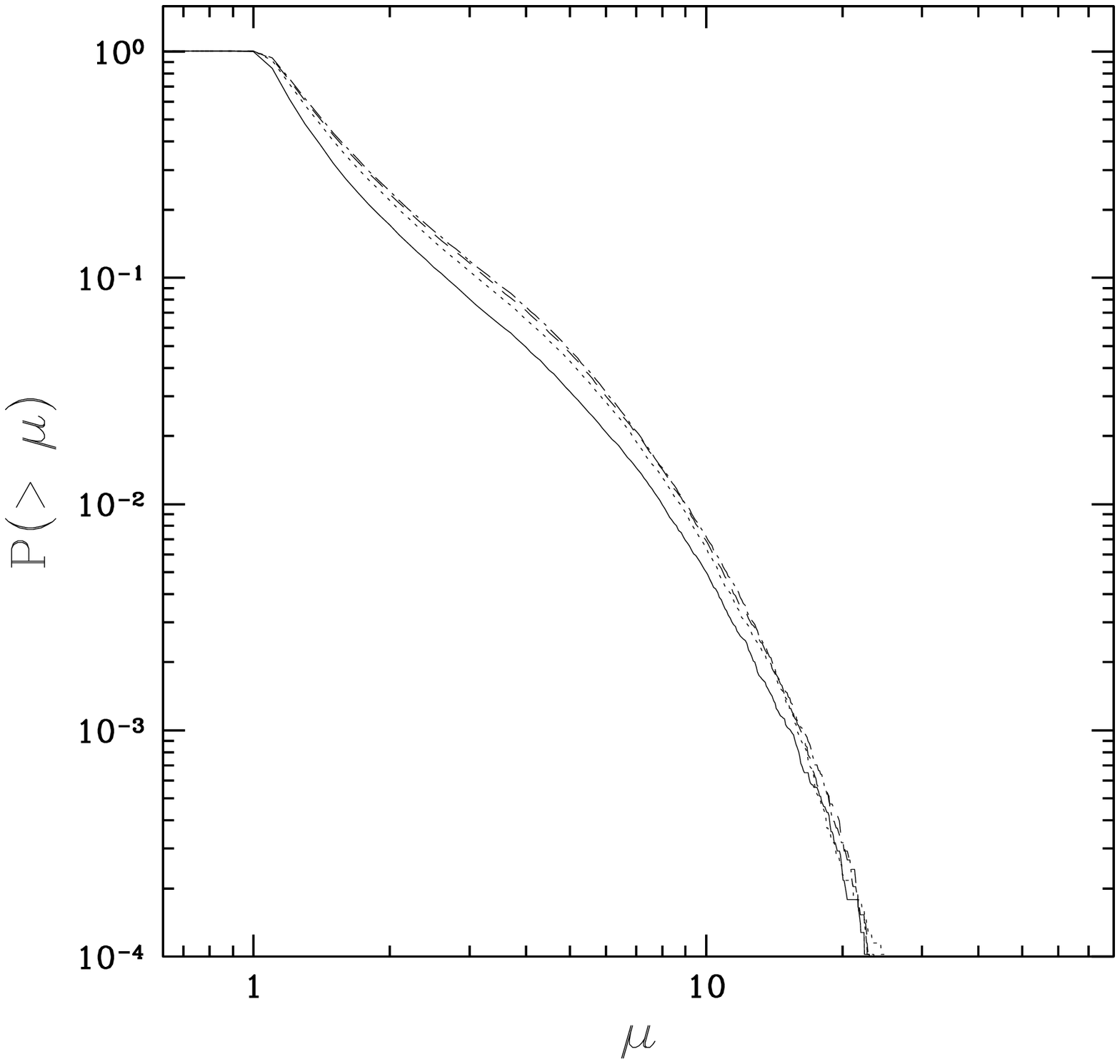}}
Fig. 4c -
\end{figure}

\begin{figure}
\centerline{\psfig{figure=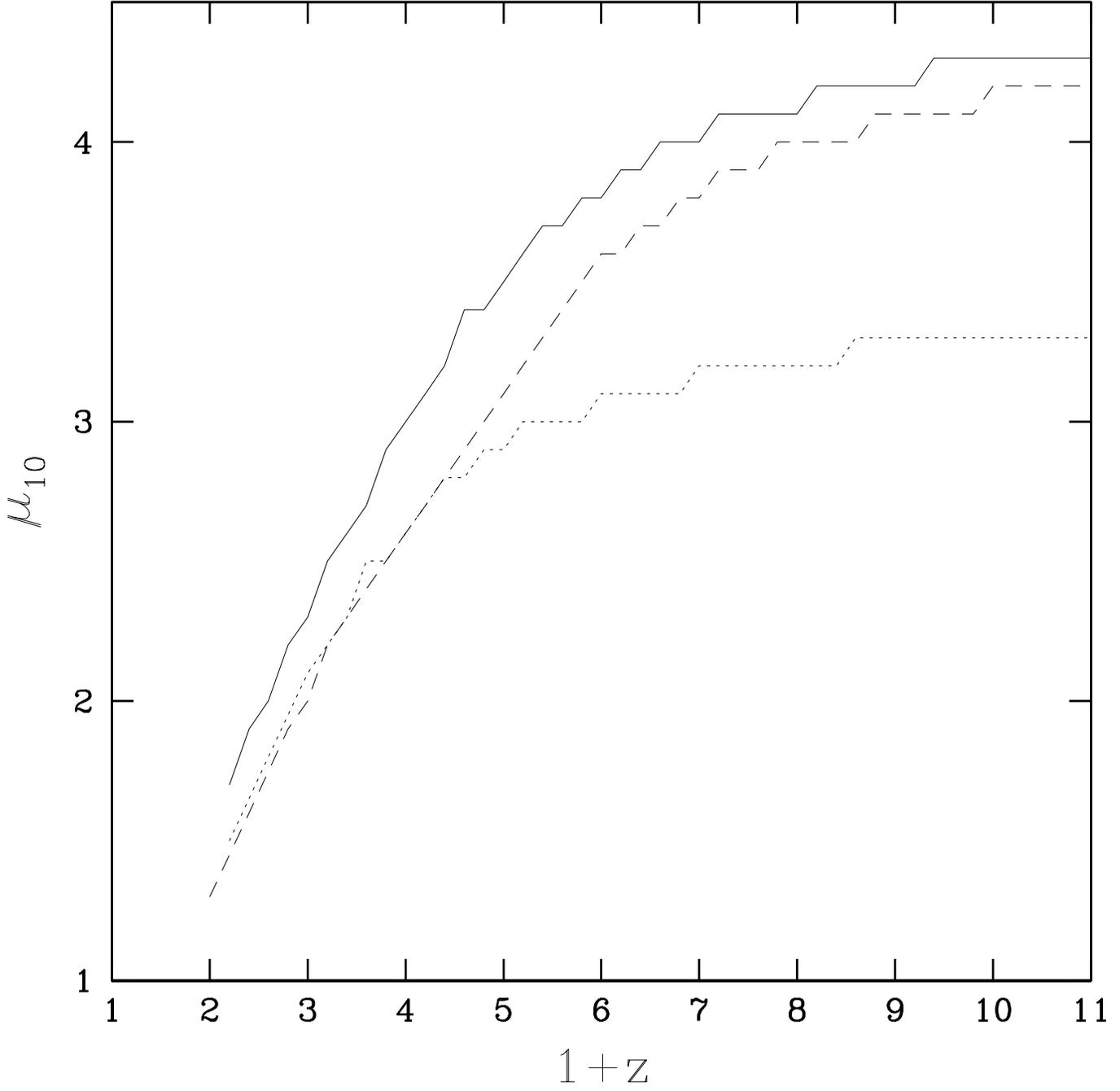}}
\caption{{\footnotesize\label{fig6} Values of magnification with associated probability 
larger than 10\% as a function of source redshift in the three cosmological 
models: SCDM (solid line), LCDM (dashed), CHDM (dotted).
 }}
\end{figure}

\begin{figure}
\centerline{\psfig{figure=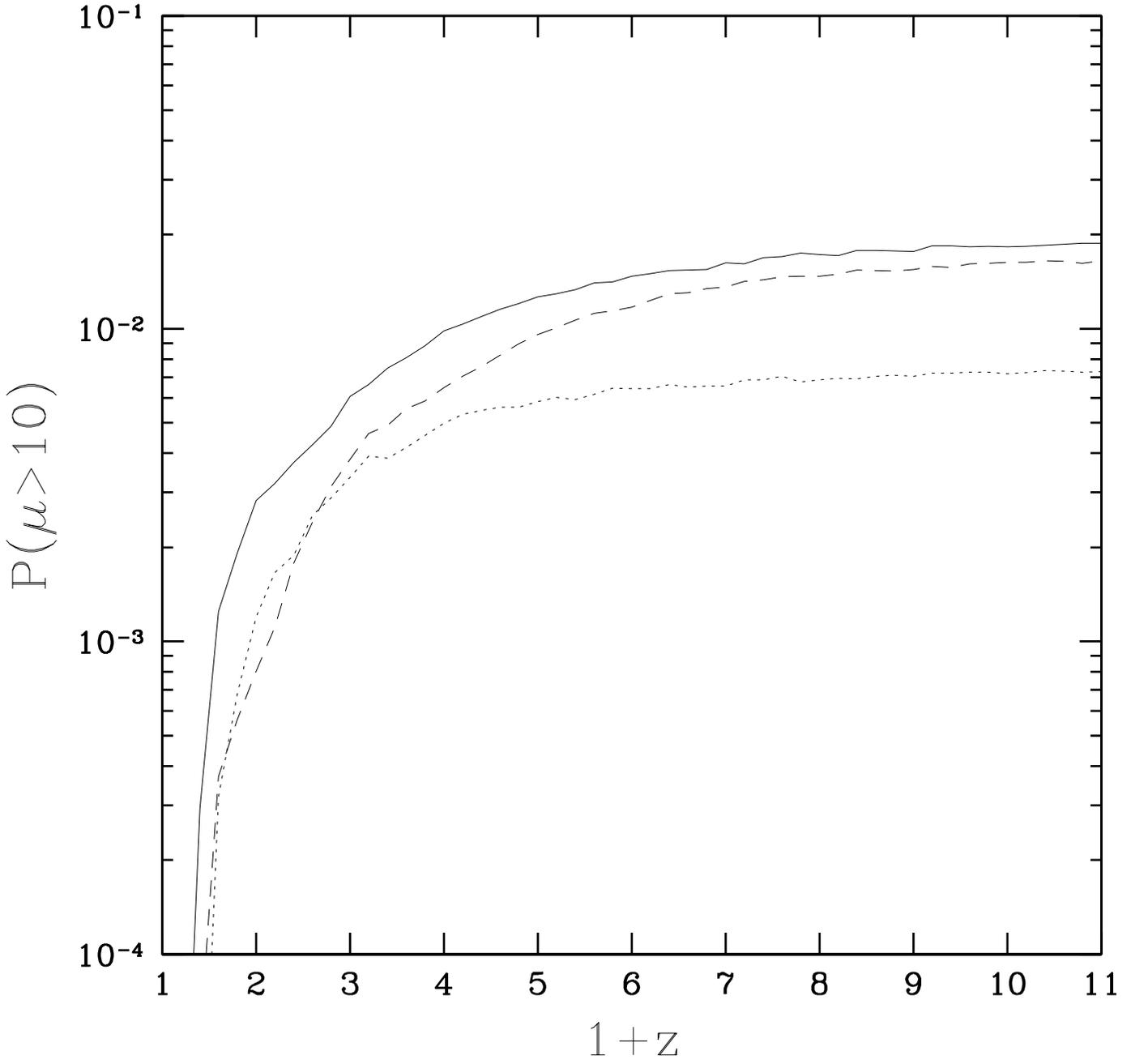}}
\caption{{\footnotesize\label{fig7} Probability associated with magnifications higher 
than $\mu=10$ as a function of source redshift in the three cosmological 
models: SCDM (solid line), LCDM (dashed), CHDM (dotted).
 }}
\end{figure}

\begin{figure}
\centerline{\psfig{figure=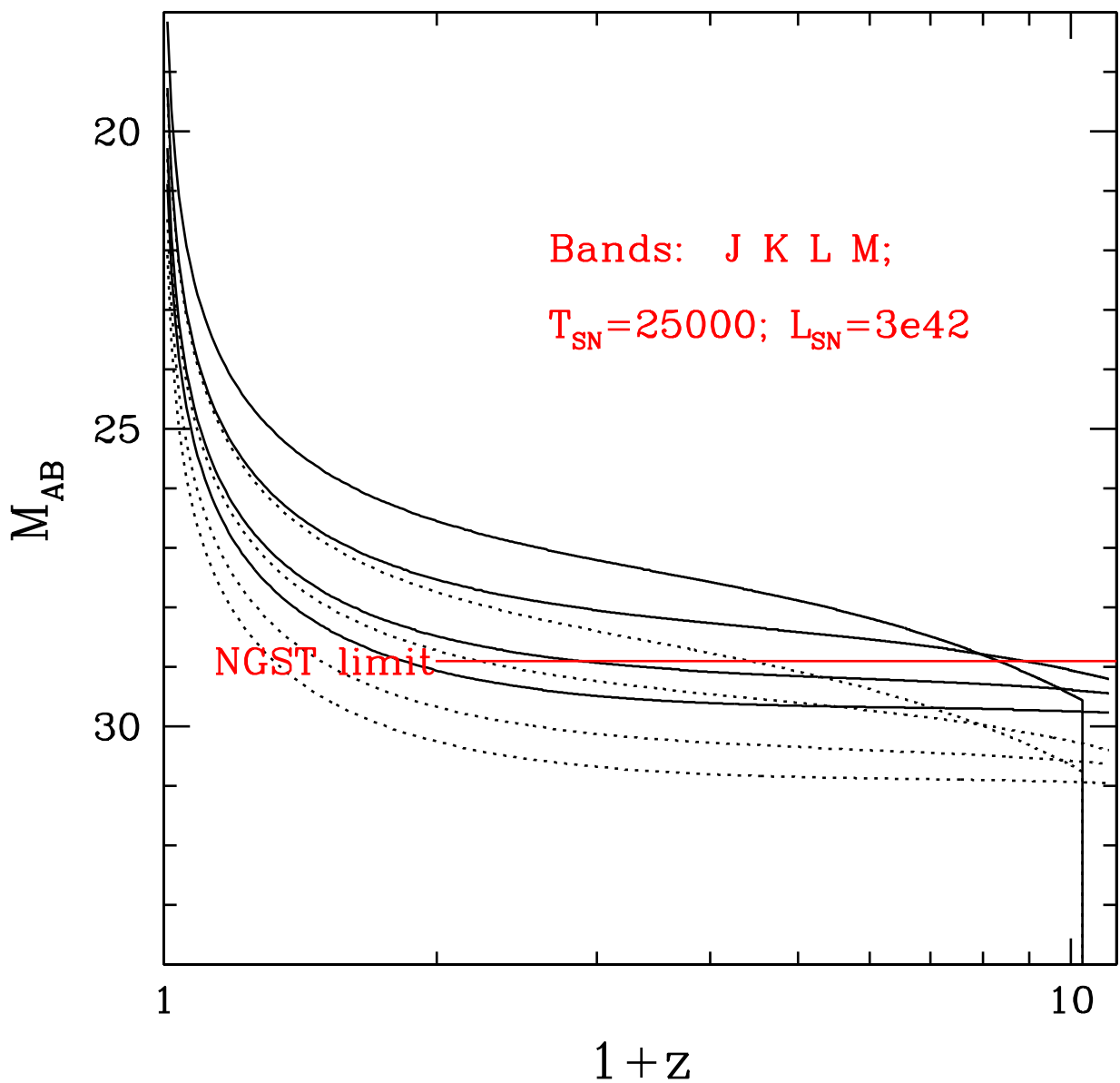}}
\caption{{\footnotesize\label{fig8} AB apparent magnitude of a SN as a function of
its explosion redshift in the four wavelength bands J, K, L, M  (from the
uppermost to the lowermost curve) for the SCDM model.
The case for a moderate ($\mu=3$) magnification (solid
curves) is compared to the one in which magnification is neglected (dotted).
The NGST flux limit is also shown; the vertical line
at high redshift in the J band is due to intergalactic absorption.
 }}
\end{figure}

\begin{figure}
\centerline{\psfig{figure=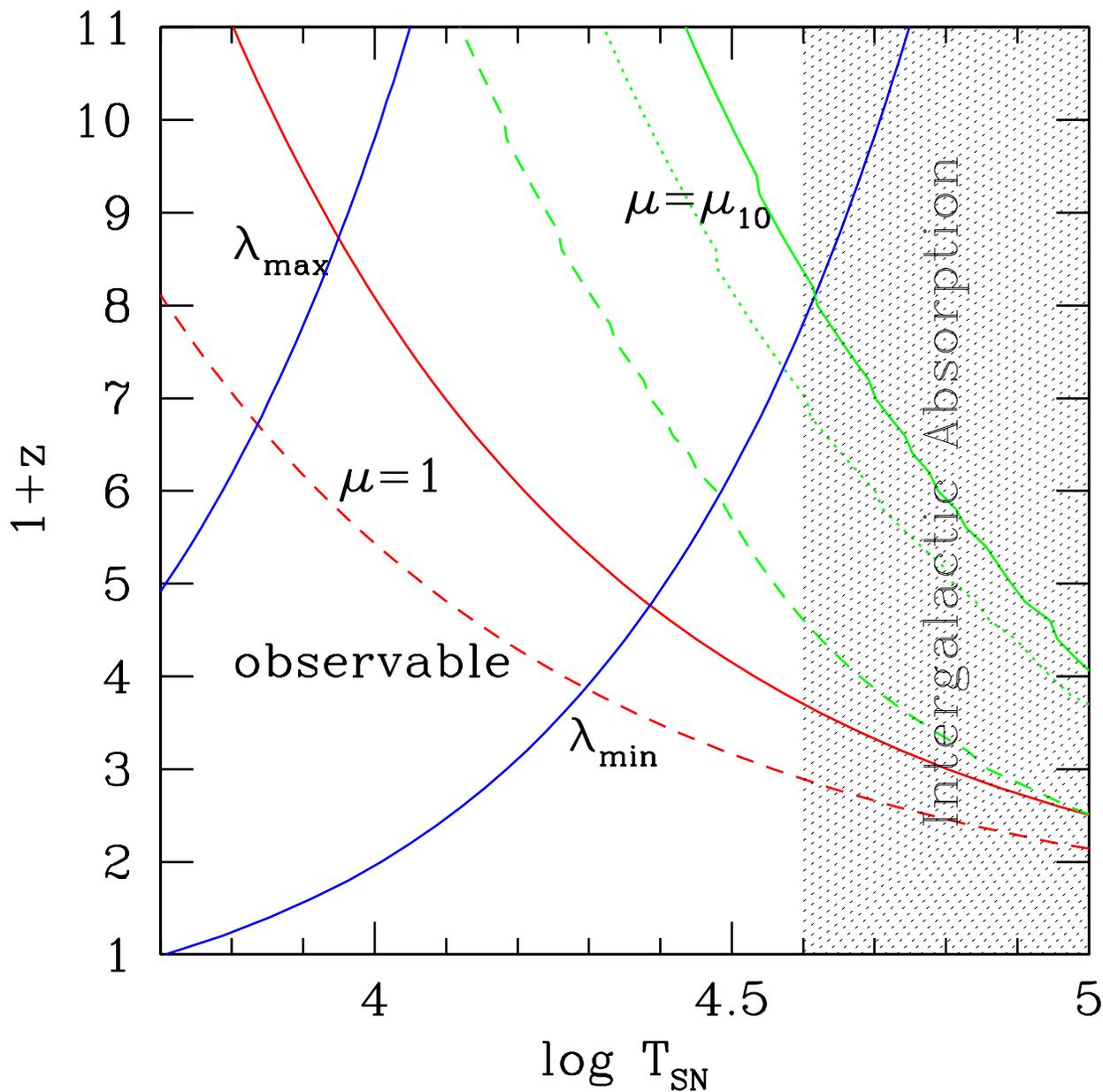}}
\caption{{\footnotesize\label{fig9} SN detection area (marked as "observable") in the
redshift-SN temperature, $T_{SN}$, plane neglecting GL magnification (curves 
$\mu=1$) and with magnification $\mu_{10}(z)$ for the three cosmological
models: SDCM (solid curve), LCDM (dashed), CHDM (dotted). Note that for $\mu=1$
the SCDM and CHDM curves overlap. The area is also
bound by the lower and upper NGST wavelength limits, $\lambda_{min}$ and 
$\lambda_{max}$, respectively. The shaded region shows the effect of
intergalactic absorption; the SN luminosity is $3\times 10^{42}$~erg~s$^{-1}$.
 }}
\end{figure}
\end{document}